\pdfoutput=1

\documentclass[sigconf]{acmart}

\usepackage{graphicx}
\graphicspath{{images/}}
\usepackage{booktabs}

\usepackage[group-separator={,},group-minimum-digits=4]{siunitx}
\usepackage{csquotes}
\usepackage[inline]{enumitem}
\usepackage{listings}
\usepackage[capitalize,noabbrev]{cleveref}

\newenvironment{boxed_new}
    {\begin{center}
    \begin{tabular}{|p{0.45\textwidth}|}
    \hline\\
    }
    { 
    \\\\\hline
    \end{tabular} 
    \end{center}
    }

\newcommand{\gh}{GitHub}
\newcommand{\pr}{pull request}
\newcommand{\tta}{time-to-accept}
\newcommand{\ttm}{time-to-merge}
\newcommand{\oss}{\textsc{OSS}}
\newcommand{\mcr}{\textsc{MCR}}

\newtheorem{observation}{\textbf{Observation}}

\acmConference[MSR 2022]{MSR '22: Proceedings of the 19th International Conference on Mining Software Repositories}{May 23–24, 2022}{Pittsburgh, PA, USA}

\begin{document}

\title{Do Small Code Changes Merge Faster?\texorpdfstring{\\}{Lg}A Multi-Language Empirical Investigation}

\begin{abstract}
\emph{Code velocity}, or the speed with which code changes are integrated into a production environment, plays a crucial role in Continuous Integration and Continuous Deployment. 
Many studies report factors influencing code velocity.
However, solutions to increase code velocity are unclear. 
Meanwhile, the industry continues to issue guidelines on \textquote{ideal} code change size, believing it increases code velocity despite lacking evidence validating the practice.
Surprisingly, this fundamental question has not been studied to date.
This study investigates the \emph{practicality of improving code velocity by optimizing \pr\ size and composition} (ratio of insertions, deletions, and modifications).

We start with a hypothesis that a moderate correlation exists between \pr\ size and \ttm.
We selected \num{100} most popular, actively developed projects from 10 programming languages on \gh.
We analyzed our dataset of \num{845316} \pr s by size, composition, and context to explore its relationship to \ttm---a proxy to measure code velocity. 
Our study shows that \pr\ size and composition do not relate to \ttm. 
Regardless of the contextual factors that can influence \pr\ size or composition (e.g., programming language), the observation holds.
Pull request data from two other platforms: Gerrit and Phabricator (\num{401790} code reviews) confirms the lack of relationship. 
This \emph{negative result} as in \textquote{\dots eliminate useless hypotheses \dots}~\cite{tichy_2000}
challenges a widespread belief by showing that \emph{small code changes do not merge faster to increase code velocity}.

\end{abstract}

\author[1]{Gunnar Kudrjavets}
\orcid{0000-0003-3730-4692}
\affiliation{
   \institution{University of Groningen}
   \city{Groningen}
   \postcode{9712 CP}
   \country{Netherlands}}
\email{g.kudrjavets@rug.nl}

\author[2]{Nachiappan Nagappan}
\orcid{0000-0003-1358-4124}
\affiliation{
    \institution{Meta Platforms, Inc.}
    \streetaddress{1 Hacker Way}
    \city{Menlo Park}
    \state{CA}
    \country{USA}
    \postcode{94025}}
\email{nnachi@fb.com}

\author[3]{Ayushi Rastogi}
\orcid{0000-0002-0939-6887}
\affiliation{
   \institution{University of Groningen}
   \city{Groningen}
   \postcode{9712 CP}
   \country{Netherlands}}
\email{a.rastogi@rug.nl}




\maketitle

\section{Introduction}
Continuous deployment is a software development methodology
that deploys a continuous stream of software updates into the
production environment~\cite{savor_continuous_2016,rossi_continuous_2016}. 
For engineers working in fast-paced environments, continuous
deployment and particularly the \emph{speed} at which code
changes are integrated into the production environment (also referred to as \emph{code velocity}) are essential~\cite{feitelson_2013}.
In these environments, since code changes constantly flow to the main branch from multiple sources,
frequently fetching from and pushing changes to the main branch prevents merge conflicts and 
ensures timely deployment.
The practice has widespread adoption across high-profile internet and social media companies like Amazon, Facebook, Google, Netflix, and Twitter.

Studies show many factors influence code velocity~\cite{yu_wait_2015,yu_determinants_2016,baysal_2016,sadowski_modern_2018,
bosu_2014,kononenko_2018,lee_2017,hilton_2016,zhao_2017,pinto_who_2018}, often referred to and measured as \tta\ and/or \ttm.
These include \emph{code characteristics} (e.g., code churn~\cite{kononenko_2018,yu_wait_2015,yu_determinants_2016}),
\emph{project characteristics} (e.g., age~\cite{yu_wait_2015,yu_determinants_2016}
and the number of open \pr s~\cite{yu_wait_2015,yu_determinants_2016}),
and \emph{human and social factors} (e.g.,
contributor affiliation~\cite{baysal_2016,sadowski_modern_2018,kononenko_2018,pinto_who_2018,baysal_influence_2013} and strength of social connection~\cite{yu_wait_2015,yu_determinants_2016}). 
Notably, only a few of these factors can be controlled by a \pr\ author to increase code velocity.

Our goal is to explore the \emph{practicality of improving code velocity by exploring solutions within the direct control of an engineer.}
For example, engineers can control the size of a proposed code change but not their reputation. 
This study investigates two such factors that can be adapted to increase code velocity: \pr\ size and composition.
To investigate whether \emph{\pr\ size and composition can be meaningfully changed to increase code velocity} (using \ttm\ as its proxy), we ask three research questions:

\begin{quote}
 \textbf{RQ1:} \emph{What characterizes \pr\ size, composition, and \ttm?}
 \end{quote}

The objective of the first research question is to give insights into the characteristics and distribution of  the data.
This understanding is crucial to identify the \emph{scope for improvements} (e.g., the extent to which \ttm\ can be decreased) and offer a \emph{preliminary understanding} (e.g., degree of variability in \pr\ size) for deeper investigations later. 

The following two questions explore direct (RQ2) and mediated (RQ3) relationships of \pr\ size and composition to \ttm\ for improving code velocity.

 \begin{quote}
 \textbf{RQ2:} \emph{What is the relationship of \pr\ size and composition to \ttm?}; and
 \end{quote}

  \begin{quote}
  \textbf{RQ3:} \emph{Does context influence the relationship of \pr\ size and composition to \ttm?}
 \end{quote}
Note that a \pr\ author cannot control contextual factors in itself; however, they can influence \pr\ size and composition, potentially influencing the ways to increase code velocity. 
 
We collected information on \num{100} \gh\ repositories: \num{10} most popular repositories each from the \num{10} most popular programming languages on GitHub.
We report descriptive statistics answering RQ1 and the correlation of \pr\ size and composition to the \ttm\ (in RQ2) in the presence of confounding factors (RQ3).
We confirm our findings by replicating our analysis on smaller datasets from code review platforms Gerrit and Phabricator.

We found that \emph{\pr\ size and its composition do not influence \ttm}, regardless of the day of the week the \pr\ is created, programming language, and organizational affiliation.
There are no patterns even though:
\begin{itemize}
\item code velocity is not the same for industry and non-industry repositories: repositories affiliated with industry have larger \pr\ sizes, but lower \ttm\ compared to non-industry repositories; and 
\item expressiveness of programming language influences both \pr\ size and \ttm. 
\end{itemize}

Challenging popular beliefs, our study suggests that neither \tta\ nor \ttm\ decreases with reduced \pr\ size
or the composition of code changes.
Splitting \pr s into smaller and isolated chunks may help with increasing
the chances of acceptance~\cite{weisgerber_small_2008} or make code
changes \textquote{reviewable}~\cite{ram_what_2018}, but it does not
decrease the overall time it takes to review or merge them.
Likewise, we do not observe that certain types of code changes merge faster than others.

\section{Background and Related Work}
\label{section:background-and-related-work}

\subsection{Introduction to Modern Code Review}
Most present-day software development organizations and open-source software (\oss) projects
have embraced a process known as Modern Code Review (\mcr)
~\cite{sadowski_modern_2018}.
The \mcr\ process starts with an engineer proposing a set of changes
(insertions, deletions, and modifications) to a source code
and submitting the collection of code changes for review.
Depending on the organization and context,
the collection of code modifications is referred to as \textquote{change},
\textquote{changelist}, \textquote{diff},
\textquote{patch} or a \textquote{\pr}. 
For example, Facebook uses \textquote{diff},
\gh\ uses \textquote{\pr s},
Google uses \textquote{changelist},
and most of the \textsc{OSS} projects use the term \textquote{patch}. 
Given that the data analysis in this paper is mainly based on GitHub,
we use the term \emph{\pr}.
After the changes are submitted for review,
the author and reviewer(s) discuss the proposed changes. 
Code review discussion may result in no critique of the original \pr,
changes being outright rejected, or some amount of code churn.
Every revision to the \pr\ will cause the review process to be
repeated until a conclusion is achieved.
If the changes are approved, the engineer can then propagate the changes to the destination branch, provided the changes pass the required automated validation tests.

A variety of tools are used to conduct \mcr.
This paper mainly uses the data mined from GitHub and compares our findings with the code review data fetched from Gerrit and Phabricator~\cite{gerrit_history,phabricator_history}.
Gerrit and Phabricator expose timestamps which enable us to determine when code
reviews are \emph{approved} (\emph{accepted}).
\gh\ only added the functionality to explicitly approve changes in 2016,
and none of the projects we mined in this study use it consistently~\cite{github_code_review_2016}.
Another conceptual difference between the tools listed above is their default
approach to merging approved changes.
The policy for Gerrit
is to start the merging process automatically.
For both \gh\ and Phabricator, manual action is required.

\subsection{Code velocity}

Studies show that a wide range of factors influences a \pr's \ttm ~\cite{yu_wait_2015,yu_determinants_2016,
baysal_2016,sadowski_modern_2018,bosu_2014,kononenko_2018,lee_2017,
hilton_2016,zhao_2017,pinto_who_2018}. 
These include personal characteristics of the author (e.g., affiliation),
project characteristics (e.g., project age),
and \pr\ characteristics (e.g., code churn).
See \cref{tab:pr-latency-factors} for a complete list of factors found to
influence the \ttm\ of projects hosted on GitHub~\cite{zhang_2020}.
We, on the contrary, explore solution space to increase code velocity by
manipulating controllable factors to reduce \ttm.

To decrease \ttm, one should be able to control and/or manipulate some or all of these factors.
By control, we mean something that an engineer proposing the change can influence in a reasonable amount of time.
Unfortunately, out of the \num{29} factors listed in \cref{tab:pr-latency-factors}, only \num{8} can be controlled by an engineer in a non-trivial manner. 

The ability to control or change a characteristic in a reasonable amount of time is important for several reasons.
First, a study of five \textsc{OSS} projects reported that more than 80\% of developers are newcomers or leavers~\cite{foucault_2015}.
This implies that most contributors may not have the time, interest, or ability to become a core member and/or increase their social strength---factors important for improving \ttm .
Second, factors such as reputation may be within an engineer's control but take months or years to change. Reputational factors are practically infeasible to optimize for a given \pr.
Third, most other remaining factors cannot be controlled by an individual at all. 
For example, project-related factor, such as the number of open \pr s, depends on various circumstances that an individual cannot easily influence.

\begin{table}[ht]
  \caption{Factors influencing \pr\ (\textsc{PR}) \ttm.}
  \label{tab:pr-latency-factors}
  \begin{tabular}{llc}
    \toprule
    Entity  & Factor & Control-\\
    &  & lable \\
    \midrule
    Author & Contributor affiliation~\cite{kononenko_2018,pinto_who_2018} & No  \\
    &Core member~\cite{yu_determinants_2016,bosu_2014,lee_2017,yu_wait_2015} &No  \\
    &First \pr~\cite{lee_2017} & No  \\
    &First response time~\cite{yu_determinants_2016,yu_wait_2015} & No  \\
    &Followers~\cite{yu_determinants_2016,yu_wait_2015} & No  \\
    &Integrator affiliation~\cite{baysal_2016,sadowski_modern_2018} & No  \\
    &Prev \pr s~\cite{kononenko_2018} & No  \\
    &Social strength~\cite{yu_determinants_2016,yu_wait_2015} & No  \\
    \midrule
    Project & Integrator availability~\cite{yu_determinants_2016,yu_wait_2015} & No  \\
    &Open \pr\ count~\cite{yu_determinants_2016,yu_wait_2015} & No  \\
    &Project age~\cite{yu_determinants_2016,yu_wait_2015} & No  \\
    & Requester success rate~\cite{yu_wait_2015} & No  \\
    &team\_size~\cite{yu_determinants_2016,yu_wait_2015} & No  \\
    \midrule
        \textsc{PR}& Continuous integration exists~\cite{hilton_2016,zhao_2017} & No  \\
    &Continuous integration latency~\cite{yu_determinants_2016,yu_wait_2015} & No  \\
    &Continuous integration test passed~\cite{yu_determinants_2016,yu_wait_2015} & No  \\
    &Commits on files touched~\cite{yu_determinants_2016,yu_wait_2015} & No  \\

    &Friday effect~\cite{yu_determinants_2016,yu_wait_2015} & No  \\
    &Number of comments~\cite{yu_determinants_2016,yu_wait_2015} & No  \\
    &Number of participants~\cite{kononenko_2018} & No  \\
    & Participants in PR/commit comments~\cite{kononenko_2018} & No  \\
    &Test inclusion~\cite{yu_determinants_2016} & Yes  \\
 & At tag~\cite{yu_determinants_2016,yu_wait_2015} & Yes  \\
    &Hash tag~\cite{yu_determinants_2016,yu_wait_2015} & Yes  \\
    &Churn addition~\cite{yu_determinants_2016,yu_wait_2015,baysal_2016} & Yes  \\
    &Churn deletion~\cite{yu_determinants_2016,yu_wait_2015,baysal_2016} & Yes  \\
    &Source code churn~\cite{kononenko_2018} & Yes  \\
        &Number of commits~\cite{yu_determinants_2016,yu_wait_2015,zhao_2017} & Yes  \\
    &Description length~\cite{yu_determinants_2016,yu_wait_2015} & Yes  \\
   \bottomrule
  \end{tabular}
\end{table}

\cref{tab:pr-latency-factors} classifies each factor influencing
\ttm\ according to our assessment of controllability. 
These factors are taken from ~\cite{gousios_14,zhang_2020}.
\cref{tab:pr-latency-factors} shows that no known author and project characteristics influencing \ttm\ can be controlled.
Some of the factors that can be controlled relate to the size and composition of a \pr\ (see \cref{subsec:choice_of_metrics}).
Obviously, attributes such as description of code changes can be optimized for grammatical errors and relevance to a change, but the semantic value of these concepts are hard to quantify.
We expect that the same factors will also influence code reviews performed using either Gerrit or Phabricator because these code reviewing platforms are semantically similar. 

\subsection{Code size}

A study focusing on the Mozilla project finds that developers \emph{feel}
that the size-related factors are the most important for code review time and decision~\cite{kononenko_code_2016}.
Another study based on interviews with 10 participants (8 from industry and
2 from \oss\ community with a median experience of $9.5$ years) finds that the
smaller the change, the more \emph{reviewable} it is,
but the precise sizes for categories such as extra small, small,
medium and large are unknown~\cite{ram_what_2018}.
In the same study, one of the interviewees states that
\textquote{for a change to be reviewable it must be at
most 250 lines long}.

A study on \gh\ finds that \pr s
with small commit sizes were more likely to be accepted~\cite{tsay_influence_2014}.
However, a large patch size by itself was not a reason for rejecting the code changes.
A study examining the patch history of Eclipse and Mozilla found that
only $0.3\%$ of the rejections were attributed to the patch size
being too large~\cite{tao_writing_2014}.
The need for the patches to be small and concise is affirmed by the
study with participants stating that \textquote{large patch sets
are difficult to review and require a lot
of time to read, thus this may delay the acceptance of the patches} and
\textquote{having small patches is very important for making it easier
to revert them}~\cite{pascarella_information_2018}.

\subsubsection{No consistent relations}
\label{subsec:related-work}
The relationship between the speed of patch acceptance and
size is not uniformly established.
In one of the earliest attempts to investigate the relationship between the
patch size and its acceptance speed, the authors found that \emph{smaller} patches
(defined as 15 lines or less for \textsc{FLAC}
and 24 lines or less for OpenAFS) have a higher chance of acceptance
than average.
However, they could not conclusively state that the patch size significantly influenced the acceptance time~\cite{weisgerber_small_2008}.
Another study researching the patch acceptance in Linux kernel found a link between patch size and time to review the changes, but not the patch integration time~\cite{jiang_will_2013}.

A study on \gh\ concludes
that \textquote{the size of a \pr\ matters: the
shorter it is, the faster it will be reviewed}~\cite{yu_wait_2015}.
A more detailed version of the same study by a subset of the original authors 
states that \textquote{the more succinct a pull-request
is, the faster it will be reviewed}~\cite{yu_determinants_2016}.
For clarity, we need to point out that though the term
\textquote{reviewed} is used, the authors of the study use
the terms \textquote{\pr\ latency} and \textquote{review latency} synonymously.
However, those two terms mark different states in a \pr\ life cycle.
Pull request latency is the difference between the \pr\ closing and
creation times. 
The proposed code changes are considered to be reviewed when another
engineer either formally or informally approves
the changes or requests additional changes that must be
implemented before the \pr\ can be merged.
Changes being reviewed do not mean that a \pr\ has been merged or closed.

Another study focusing on a single \gh\ project (Shopify's
Active Merchant) shows that the size of a \pr\ had a statistically
significant effect on review time~\cite{kononenko_2018}.
The study consisted of a dataset of \num{1475} \pr s.
Another study of \num{97463} \pr s from \num{30} projects
suggests a relationship between the number of source lines of code (\textsc{SLOC})
modified and \pr\ lifetime but concludes that there are no means to establish
that an increase in \textsc{SLOC} implies a longer \pr\ lifetime ~\cite{soares_2020}.

\subsubsection{Size-related recommendations}
Reflecting on the assumption that size matters to \textsc{MCR},
industry and \textsc{OSS} projects recommend that \pr\ size be \textquote{small}.
The guidance, however, is general and vague.
Projects suggest \emph{isolated} changes of \emph{reasonable} size.
Popular \textsc{OSS} projects such as
\begin{enumerate*}[label=(\roman*),before=\unskip{ }, itemjoin={{, }}, itemjoin*={{, and }}]
    \item Linux directs contributors to \textquote{Separate each logical change into a separate patch} and suggests that
    \textquote{It cannot be bigger than 100 lines \dots}
    ~\cite{linux_patch_guidance,linux_no_more_than_100}
    \item LLVM recommends that the patch should \textquote{be an isolated change}~\cite{llvm_patch_guidance} 
    \item Chromium contribution guidelines say that \textquote{Patches should be a reasonable size to review} ~\cite{chromium_patch_guidance}
    \item PostgreSQL patch submission guidance suggests \textquote{Start with
    submitting a patch that is small and uncontroversial}~\cite{postgresql_patch_guidance}.
\end{enumerate*}

What exactly is a \emph{small} size to submit for a \pr? 
Challenging the perceptions of developers, a study shows that what engineers think about the size of commits differs from reality by more than
an \emph{order of magnitude}~\cite{riehle_developer_2012}. 
Even if we consider \emph{small} as a recommended size, the suggested range differs depending on the source.
A study on commits in \textsc{GNU}
Compiler Collection codebase classifies the changes based on
\textsc{SLOC} into the following categories:
\begin{enumerate*}[label=(\roman*),before=\unskip{ }, itemjoin={{, }}, itemjoin*={{, and }}]
    \item extra-small (0--5)
    \item small (6--46)
    \item medium (47--106)
    \item large (107--166)
    \item extra-large (167--\num{203359})~\cite{alali_2008}.
\end{enumerate*}

Guidelines from industry, while also emphasizing small and
incremental changes, are likewise vague.
\begin{enumerate*}[label=(\roman*),before=\unskip{ }, itemjoin={{, }}, itemjoin*={{, and }}]
    \item Google engineering practices suggest that \num{100} lines are a reasonable
    size, and \num{1000} lines are considered too large~\cite{google_eng_practices}
    \item Phabricator, the code review platform utilized at Facebook, uses
    \textquote{Each commit should be as small as possible, but no smaller}
    as guidance~\cite{phabricator_patch_guidance}
    \item Microsoft's recommended practices state that \textquote{Authors
    should aim for small, incremental changes that are easier to understand}
    ~\cite{macleod_code_2018}.
\end{enumerate*}

\subsubsection{Empirical data on code review size}
\label{subsection:emp-code-size-findings}
A study on code review practices at Google reported that the median number
of lines modified is 24~\cite{sadowski_modern_2018}. 
At Facebook, each deployed software update (code change
reviewed, committed, and then deployed to production) is
on average 92 lines~\cite{rossi_continuous_2016}.
Another investigation covering both industrial and \textsc{OSS} projects
found that the median change size for Android and AMD is 44 lines;
Apache is 25 lines; Linux is 32 lines; and Chrome is 78 lines
~\cite{rigby_convergent_2013}. 
However, the same study also found that for Lucent, the change size was an order of magnitude bigger, with the median being 263 lines.
A study exploring the review-then-commit type of changes reported
median churn between 11 and 32 lines~\cite{rigby_peer_2014}.
One of the first studies that investigated the relationship between
patch size and its acceptance time in two \textsc{OSS} projects
(\textsc{FLAC} and OpenAFS) noted that for \textsc{FLAC}
more than half of the submitted patches change one or two lines of code;
for OpenAFS, one-third of the patches change at most two lines
~\cite{weisgerber_small_2008}. 
Finally, analysis of many \gh\ repositories shows that the median number
of lines changed by \pr s is 20~\cite{gousios_exploratory_2014}.

\begin{observation}
\label{obs:median_code_change_differs}
Both industrial and \textsc{OSS} projects recommend \textquote{small}
or \textquote{reasonable} sizes for \pr s, but the guidance is vague.
Empirical data shows that actual median code change size differs by orders of magnitude across projects in practice.
The lack of concrete guidance may be a driving factor for the
wide variance.
\end{observation}

\subsection{Code composition}

When referring to code change size, most studies define code churn as the sum of added and deleted lines~\cite{yu_determinants_2016,
baysal_influence_2013,yu_wait_2015,weisgerber_small_2008,
jiang_will_2013,ram_what_2018}.
Only two studies classify code changes in detail.
One study uses the term \emph{change} to describe technical contributions
with no specific definition~\cite{tsay_influence_2014}. 
Another study investigating the characteristics of commit sizes
separates change types into \emph{add}, \emph{modify}, and \emph{delete}~\cite{alali_2008}.

Most of the data gathered in the above studies are produced by the various diff analysis tools or
fetched from GitHub.
\gh\ uses the output of the \verb|git diff| command (addition and deletion) as one of the
attributes describing a \pr.
Unfortunately, using only insertions and deletions to estimate
the code churn results in an erroneous estimate of the amount of code changed
because moves and updates are not accounted for~\cite{canfora_identifying_2007}.

\section{Study Design}
\label{section:study-design}

\subsection{Choice of data}
We sought data from \gh\ repositories that are popular, actively developed, and cover a wide range of programming
languages.
Our dataset is available here\footnote{https://figshare.com/s/37af0f90d3d2a15f1762} for replication.
In the final version, we will also share the scripts, which as of now, contain personally identifiable information.

By analyzing the most popular and actively developed repositories we gain many benefits: 
\begin{enumerate*}[label=(\roman*),before=\unskip{ }, itemjoin={{, }}, itemjoin*={{, and }}]
    \item a larger dataset populated by a continuous flow of incoming \pr s by a variety of contributors
    \item a larger number of core developers with permission to approve merges (thus eliminating artificial bottlenecks in the \pr\ review process)
    \item an accurate representation of the dynamics of collaboration on \gh\ among a diverse set of individuals.
\end{enumerate*}

To obtain \gh\ data, we considered the
following approaches:
\begin{enumerate*}[label=(\roman*),before=\unskip{ }, itemjoin={{, }}, itemjoin*={{, and }}]
    \item GHTorrent dataset and tool suite~\cite{gousios_ghtorrent_2013}
    \item GH Archive~\cite{gh_archive_main_page}
    \item GitHub API~\cite{github_rest_api_docs}.
\end{enumerate*}
We chose the \gh\ \textsc{API} to fetch up-to-date information about the state of \gh\ directly.
\gh\ \textsc{API} exposes a set of functionality that enables callers to
query and search public repositories and fetch various entities
associated with them (e.g., commits, \pr s, and users).
The other methods provide only historical snapshots.
Utilizing the \gh\ API gave us the most flexibility in
determining how and what data to collect.

A collection of code review data from various Gerrit projects is
available for researchers~\cite{yang_2016}.
The dataset is in the form of a {MySQL} database and as of July 2nd of 2021,
it contains code review data about Eclipse, GerritHub, LibreOffice, and OpenStack.
Several major \textsc{OSS} projects also use Phabricator to perform
code reviews.
We utilize Phabry to mine the publicly accessible code review data for
Blender, FreeBSD, \textsc{LLVM}, and Mozilla~\cite{cotet_2019,phabry_2021}.

\subsection{Selection and elimination criteria}

To note popularity, \gh\ uses the concept of \emph{stars}.
This idea is similar to \emph{likes} used in social media
networks like Facebook, Instagram, and Twitter.
Another paradigm used is \emph{forks}, which allows a user
to create their own copy of a repository without affecting the original
repository. Forks, stars, and the number of \pr s in a particular repository
have all been used as criteria for selecting \gh\ projects
~\cite{hilton_2016,soares_2020,borges_understanding_2016}.

Existing studies investigating factors that impact the popularity
of \gh\ repositories have found a strong positive
correlation between stars and forks, making both suitable proxies to
measure popularity~\cite{borges_understanding_2016,papamichail_2018}.
Stars, however, have multiple uses in \gh: as bookmarks, as a way to show
appreciation to repository contributors for their work, and as a way
to improve \gh\ recommendations for similar projects.
Contrastingly, when a developer forks their copy of a repository,
this act signifies an intent to modify the original code.
We find the number of forks as a better selection criterion for active development in a particular repository.

We used \gh 's report ranking
the 10 most popular programming languages for October 2019 through
September 2020~\cite{the_state_of_the_octoverse} to select programming languages.
The top 10 languages were: C, C++, C\#, Java, JavaScript,
PHP, Python, Ruby, Shell, and TypeScript.
Our collection covers object-oriented, procedural, and scripting languages.

\subsection{Choice of metrics}
\label{subsec:choice_of_metrics}

\subsubsection{Code size}

The code size is expressed as the number of commits,
files, or \textsc{SLOC} included in a \pr.
We choose to quantify the size in \textsc{SLOC} because it is a metric easily
comprehended, has support in the Gerrit, \gh, and Phabricator
infrastructure, and is already used as a primary numeric value to guide
the size of code contributions for various projects~\cite{google_eng_practices,linux_no_more_than_100,ram_what_2018}.

\subsubsection{Code composition}
We added \emph{modified} lines as a separate category from insertions and deletions to improve the granularity with which code churn is measured.
Modifications are conceptually different from insertions and deletions.
With the current metric, one modification is erroneously reported as one insertion and one deletion (e.g., a trivial example of removing an extra semicolon from the end of the line).
We believe that differentiating the types of code changes will enable greater insight into how different types of code changes impact \ttm.
To analyze code changes in each \pr\
and extract the number of modifications of each type,
we use the \verb|diffstat| tool.

Below is an example showing the difference between the two approaches for
a minor bug fix from the OpenSSL codebase~\cite{openssl_code_change}.
See code \cref{code:openssl}.

\lstset{language=C,breaklines=true,basicstyle=\ttfamily\tiny,caption={Sample OpenSSL code snippet.},label=code:openssl}
\begin{lstlisting}
diff --git a/crypto/evp/m_sigver.c b/crypto/evp/m_sigver.c
index bdcac90078..57c8ce78a4 100644
--- a/crypto/evp/m_sigver.c
+++ b/crypto/evp/m_sigver.c
@@ -60,7 +60,7 @@ static int do_sigver_init(EVP_MD_CTX *ctx, EVP_PKEY_CTX **pctx,
     }

     if (ctx->pctx == NULL) {
-        if (libctx != NULL)
+        if (e == NULL)
             ctx->pctx = EVP_PKEY_CTX_new_from_pkey(libctx, pkey, props);
         else
             ctx->pctx = EVP_PKEY_CTX_new(pkey, e);
\end{lstlisting}

When using:

\lstset{language=Bash,,breaklines=true,basicstyle=\ttfamily\small,caption={},label=}
\begin{lstlisting}
git show 5b888e931b64a132a --shortstat
\end{lstlisting}

the result is 

\begin{lstlisting}
1 file changed, 1 insertion(+), 1 deletion(-)
\end{lstlisting}

as opposed to the output from \texttt{diffstat -Cm}

\begin{lstlisting}
1 file changed, 1 modification(!)
\end{lstlisting}

The difference in estimating the total changes,
even for a trivial example above, is two times.
This describes and quantifies the intent behind the choice of code composition metric with
greater accuracy.
Still, we cannot distinguish actions such as moving chunks of source code from one location to another. 
This paper uses insertions, deletions, and modifications only.

\subsubsection{Code velocity}
Modern code review systems (such as CodeFlow, Critique, Gerrit, and Phabricator) typically track the amount of time it takes for
code changes to be \emph{accepted} or \emph{signed off} (i.e., someone other
than the author formally decides that the proposed changes can be merged either
in their original form or with some modifications).
We refer to that point in time as the \emph{\tta}.
However, for our study, we believe that a more relevant metric is the time it takes for code changes to deploy.
Code changes are not \textquote{real} until they have been merged
into a destination branch.
The change will be available for building, profiling, testing, and execution in the production or test environments only after it has been merged into the main branch. 

We use the term \emph{\ttm}\ to indicate \textquote{\ldots the time
since the proposal of a change (\ldots) to the merging in the
codebase \ldots})~\cite{izquierdo-cortazar_2017}.
Terms like \textquote{\pr\ latency} and \textquote{\pr\ lifetime} have
been used synonymously to describe the same concept.
However, we will use the term \textquote{\ttm} as we believe it precisely conveys our intent.

Other reasons justifying our focus on \ttm\ instead of \tta\ are:
\begin{enumerate*}[label=(\roman*),before=\unskip{ }, itemjoin={{, }}, itemjoin*={{, and }}]
    \item even after the formal acceptance of code changes, a non-trivial amount of time may be spent on getting the changes ready to be merged (e.g., applying the code review feedback, resolving merge conflicts, repeating some of the initial validation, and investigating test case failures in an extended test suite which were not executed during early validation.)
    \item only a few projects on GitHub formally track when the proposed changes were accepted, leaving us, therefore, with a limited dataset for analysis
    \item industry experience indicates that \ttm\ is a critical factor to
    measure. A study about Xen Project's (hypervisor software) code review experience states that
    \textquote{Xen agreed that [\ttm] was the most important parameter to track when
    considering delays imposed by the review process}
    ~\cite{izquierdo-cortazar_2017}.
\end{enumerate*}

\begin{observation}
\label{obs:ttm_matters_not_acceptance_time}
The \ttm\ is a more accurate descriptor for code changes being included
into a repository than code review acceptance time.
\end{observation}

To identify when a particular \pr\ was created, we used the \texttt{created\_at}
property of \pr s.
To calculate the \ttm, we considered two \pr\ attributes available through
\gh\ API: \texttt{closed\_at} and \texttt{merged\_at}.
Semantically, \texttt{merged\_at} is the better value because
we care about when the physical code changes are merged versus
when the \pr\ is marked as closed.
The difference between \texttt{merged\_at} and \texttt{created\_at}
represents the time it took for a \pr\ to be merged.

\subsection{Data extraction}
\label{subsection:data-extraction}

We deployed a custom tool in C\# that uses a \gh\ \textsc{API} client
library for \textsc{.NET} to gather data about the most forked
repositories for each programming language~\cite{github_octokit}.
For each programming language, we retrieved a list of the 100 most
forked projects and the number of merged \pr s per project.
We ordered the list by the number of merged \pr s in descending order.
After manual inspection and elimination of projects not directly related to developing software (e.g., code samples, coding interview study guide, solutions to the programming problems,  and storage), we picked \num{10} projects per language with the most merged \pr s.
Finally, we used \verb|curl| to fetch the contents of code changes included in each
\pr\ and parsed the resulting data locally using \verb|diffstat|~\cite{diffstat}.

We obtained \num{845316} \pr s in our
raw dataset for analysis.
After removing entries
that did not have any code changes (e.g., changes to the binary files and generally
anything not considered text by the diff tools), we ended up with \num{842303}
\pr s.
We removed one \pr\ with inconsistent timestamp (the time of the
merge was set to earlier than its creation time).
We also removed all entries that appeared to have zero \ttm\
to ensure that the dataset reflects \pr s that involve meaningful review.
Our final sample set contains \num{826259} \pr s.

\subsection{Data preprocessing}
\label{subsubsec:treatment_of_outliers}

During the manual inspection of \pr s, we noticed that several \pr s contain millions of \textsc{SLOC}.
To investigate the impact of large \pr s on total code churn, we sorted the \pr s in the descending order of the \textsc{SLOC} modified.
We then calculated what percentage of total code
churn the outlier values are responsible for.
Our analysis shows that
\num{100} \pr s ($0.01\%$ of total) are responsible for $23\%$;
\num{1000} ($0.12\%$ of total)  for $53\%$; and
\num{10000} ($1.21\%$ of total) for $76\%$ of total code churn.

Initial sampling and manual review indicated that these outlier \pr s represent mostly classical
merges between branches
and, for our study, are not representative
of the types of \pr s we want to investigate.
We applied Tukey $1.5\times IQR$ fence
exclusion criteria to identify outliers to understand their characteristics and whether the merges
are a majority of the outliers ~\cite{tukey_1981}.
We then selected 100 random \pr s
from the entries we would potentially remove to avoid distortion of our
dataset and categorized them further.
The results from the manual classification are shown in ~\cref{tab:sloc-outlier-exclusion-reason}.

\begin{table}[ht]
  \caption{Distribution of outliers excluded based on \textsc{SLOC}. Random sample of \textit{N} = 100.}
  \label{tab:sloc-outlier-exclusion-reason}
  \begin{tabular}{lr}
    \toprule
    \textbf{Reason}  & \textbf{Count}\\
    \midrule
      New code (feature, scenario) & 31\\
      Major refactoring (move, rename) & 25\\
      Backporting & 12\\
      Dependency update & 8\\
      Bug fix & 7\\
      Documentation update & 7\\
      Merge between branches & 6\\
      Dead code removal & 4\\
   \bottomrule
  \end{tabular}
\end{table}

In our random sample, most outliers are
new code, major refactoring, or backporting existing code.
Therefore, we decided not to trim the initial dataset by removing the potential outliers.
To ensure the validity of our conclusions, we analyzed the data both with and without the outliers. 
Without the outliers, the size of our dataset is \num{613007} \pr s.

\subsection{Statistical analysis}
\label{section:rq}
To answer RQ1, we report descriptive statistics and visualizations indicating the distribution of \pr\ size, composition, and \ttm\ in our data.
We report statistically significant results at a $p < .001$ and use \textsc{APA} conventions~\cite{apa}.
For RQ2, we compute Spearman rank correlation coefficients~\cite{spearman} of \pr\ size and composition to \ttm.
We choose a non-parametric measurement because 
Shapiro-Wilk tests~\cite{shapiro} show that neither the \pr\ size ($W = 0.031, p < .001$) nor \ttm\ ($W = 0.12, p < .001$) are normally distributed.
In addition, Spearman correlation is considered to be robust to outlier values~\cite{spearman_robust_outliers}.

For RQ3, we examine the influence of three contextual factors: choice of programming language, affiliation (industry versus non-industry), and the day of the week the \pr\ was created.
These factors can potentially influence \pr\ size and composition, and hence the \ttm.
For example, existing studies show that the amount of \textsc{SLOC} needed
to solve the same problem varies greatly depending upon which programming language
is used~\cite{nanz_2015, prechelt_empirical_2000}.
A study
finds that \textquote{Java programs are on average 2.2--2.9 times longer than
programs in functional and scripting languages}~\cite{nanz_2015}.
Given that we are investigating the relationship between
\textsc{SLOC} and \ttm, it is essential to differentiate between the
various groups of programming languages.

We investigate the influence of contextual factors on \pr\ size and composition.
In the case of differences in distribution, we investigate whether it influences the relations of \pr\ size and composition to \ttm.
We repeat our analysis on Gerrit and Phabricator (as is feasible), with the addition of \tta.

\section{Results}
\label{section-results}

\subsection*{RQ1: What characterizes \pr\ size, composition, and \ttm?}

Earlier, in \cref{subsection:data-extraction},
we described the method and reasoning behind not removing the outliers.
We observe that the presence of outliers has a significant impact
on the measures of central tendency.
Median values differ by approximately two
times, and the mean differs by orders of magnitude
(see ~\cref{tab:before-and-after-outlier-data}).

\begin{table}[ht]
  \caption{Time-to-merge (hours) and \textsc{SLOC} per \pr\ before and after removing the outliers using Tukey $1.5\times IQR$ fence. N = total number, M = mean, Mdn = median, SD = standard deviation.}
  \label{tab:before-and-after-outlier-data}
  \begin{tabular}{lrrrrrrr}
    \toprule
    &&\multicolumn{3}{c}{Time-to-merge}&
    \multicolumn{3}{c}{SLOC}\\
    \cmidrule(lr){3-5}
    \cmidrule(lr){6-8}
    Type&\textit{N}&\textit{M}&\textit{Mdn}&\textit{SD}&\textit{M}&\textit{Mdn}&\textit{SD}\\
    \midrule
    Before & \num{826259}& 182& 17&  860& 584& 22& \num{17123}\\
    After & \num{613007}& 28& 9& 43& 32& 12& 44\\
    \bottomrule
  \end{tabular}
\end{table}

\begin{figure}[!htbp]
    \centering
    \includegraphics[width=0.4\textwidth,keepaspectratio]{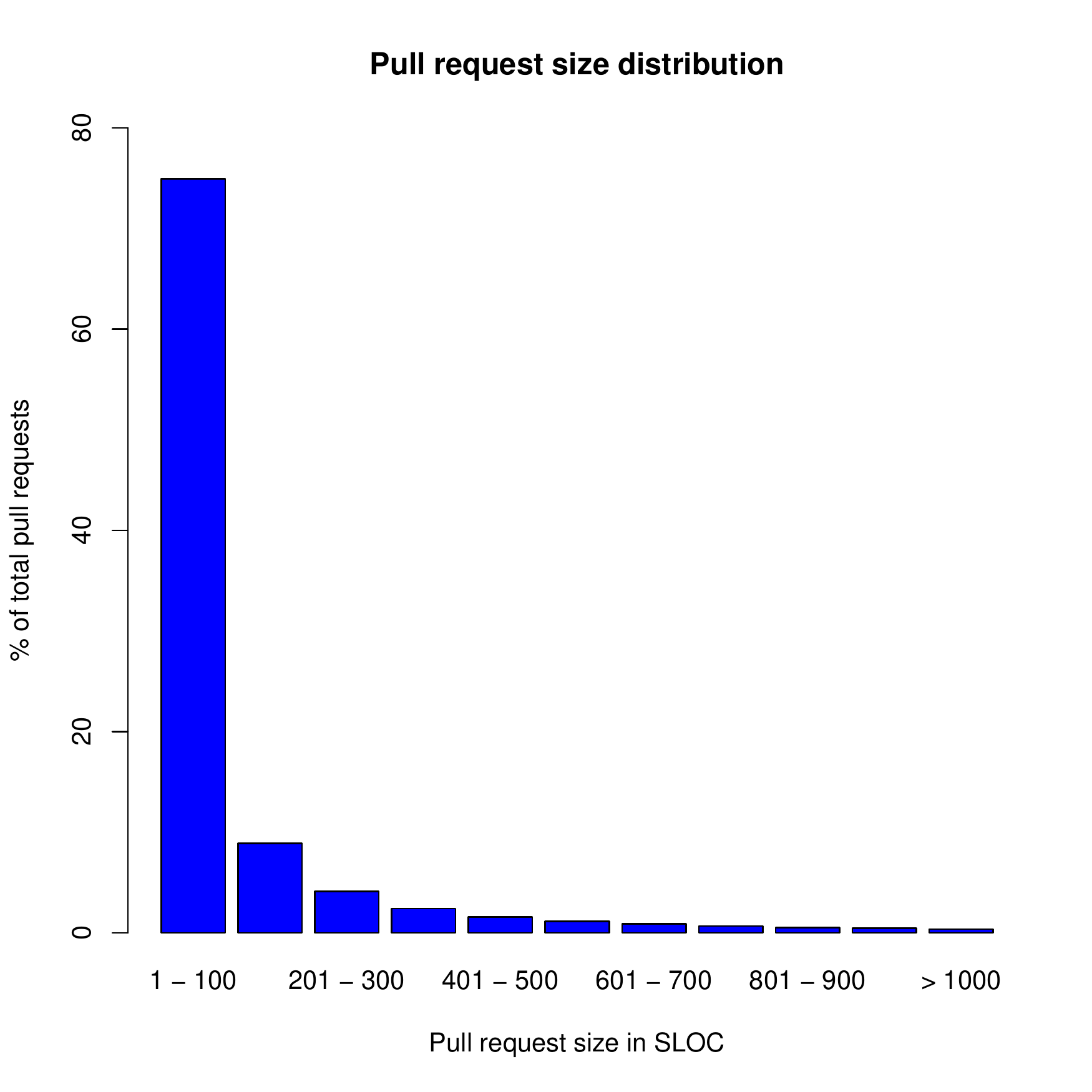}
    \caption{A barplot of \pr\ sizes.}
    \label{fig:barplot_pr_sloc}
\end{figure}

\begin{figure}[!htbp]
    \centering
    \includegraphics[width=0.4\textwidth,keepaspectratio]{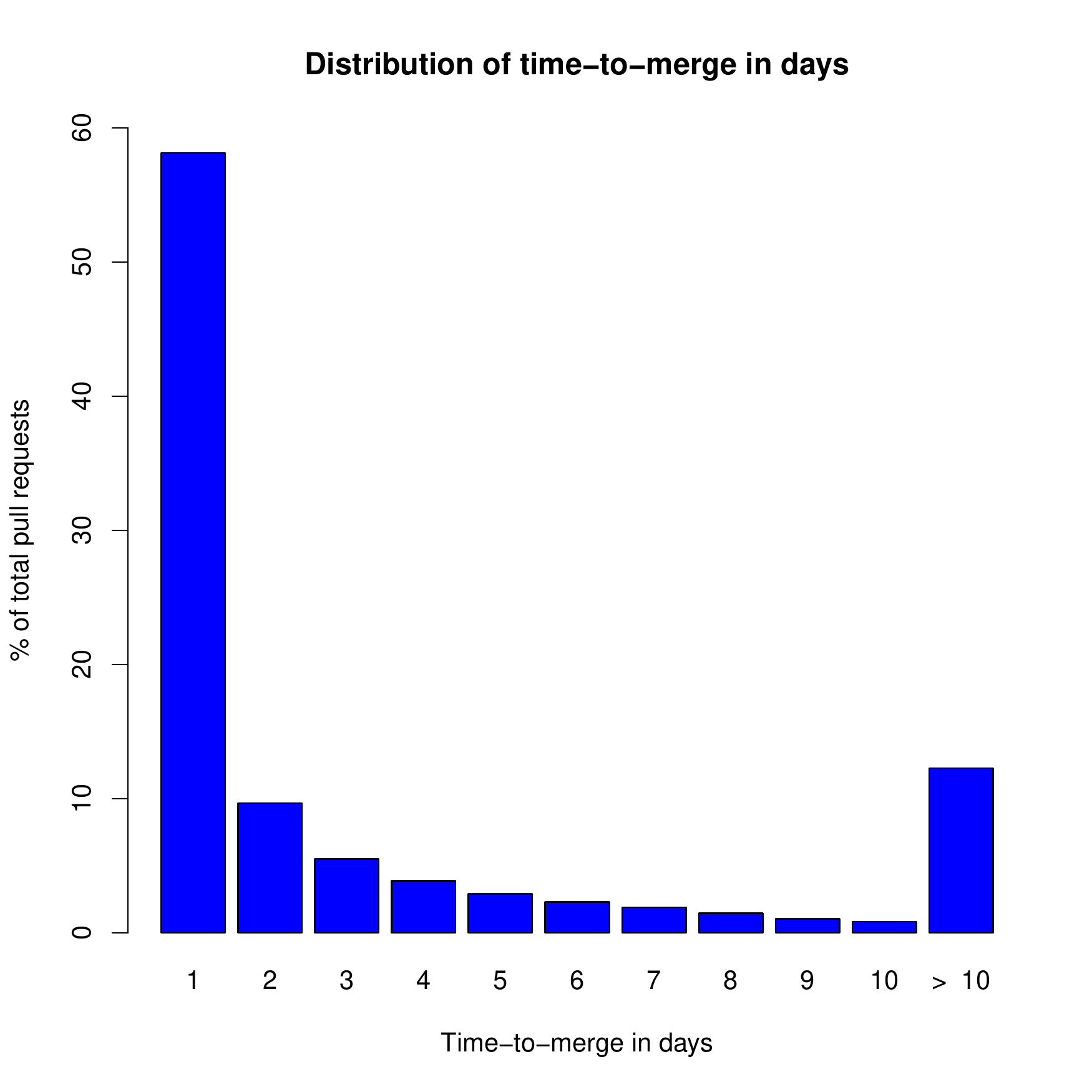}
    \caption{A barplot of \ttm\ for \pr s.}
    \label{fig:barplot_pr_ttm}
\end{figure}

Given the prevailing guidelines and beliefs surrounding the relationship
between the size of a proposed change and the \ttm, we expected
that both the \pr\ size and \ttm\  distribution would be
skewed towards the smaller values.
As expected, \cref{fig:barplot_pr_sloc} and \cref{fig:barplot_pr_ttm}
show the size of the \pr\ and \ttm\ clusters towards smaller values.
Both distributions are visually heavily right-skewed.

This finding is consistent
with what has been observed in the industry.
A study that mined data from 9 million Google code
reviews determined that modifications involving a single line
account for $>10\%$ of changes~\cite{sadowski_modern_2018}.
In our dataset, single line modifications account for $>11\%$
of changes, and unsurprisingly, the mode value for
\textsc{SLOC} changed is 1.
Our observation also closely matches what another study has observed
after inspecting the code change patterns in a codebase of
2 million \textsc{SLOC} over a decade: \textquote{nearly 10 percent
of all the changes made during the maintenance of a software under
consideration are one-line changes}~\cite{purushothaman_toward_2005}.

This finding raises a question about the link between the
guidance and the size of code changes we observe in our dataset.
Are the changes small because engineers follow the community
guidance about \pr\ size, or are they small due to the distribution
and nature of changes made?
For example, does our dataset reflect more bug fixes that tend
to be smaller versus larger modifications such as
implementing a new feature?
Unfortunately, most of the repositories we included in our study
do not formally categorize their \pr s based on the issue type
(e.g., code defect, feature, refactoring), and we do not have
enough data to answer this question.

Existing literature about the distribution of insertions,
deletions, and modifications in various codebases is limited.
Most studies use only insertions and deletions as code type changes.
Existing anecdotal industry experience
prompts us to expect engineers to insert
or delete more code than they modify.
Our dataset shows a similar trend.

\begin{boxed_new}
\textbf{RQ1:} \emph{Most \pr s are relatively small
($50\% \leq 22$ \textsc{SLOC},
$75\% \leq 101$ \textsc{SLOC}, and
$90\% \leq 381$ \textsc{SLOC}).
Half of the \pr s get merged in a matter of hours
($50\% \leq 18$ hrs,
$75\% \leq 82$ hrs, and
$90\% \leq 318$ hrs).
Insertions outnumber modifications by a $3:1$ ratio.
Deletions outnumber modifications by a $3:2$ ratio.
}
\end{boxed_new}

\subsection*{RQ2: What is the relationship of \pr\ size and composition to \ttm?}

\begin{figure}[!htbp]
    \centering
    \includegraphics[width=0.4\textwidth,keepaspectratio]{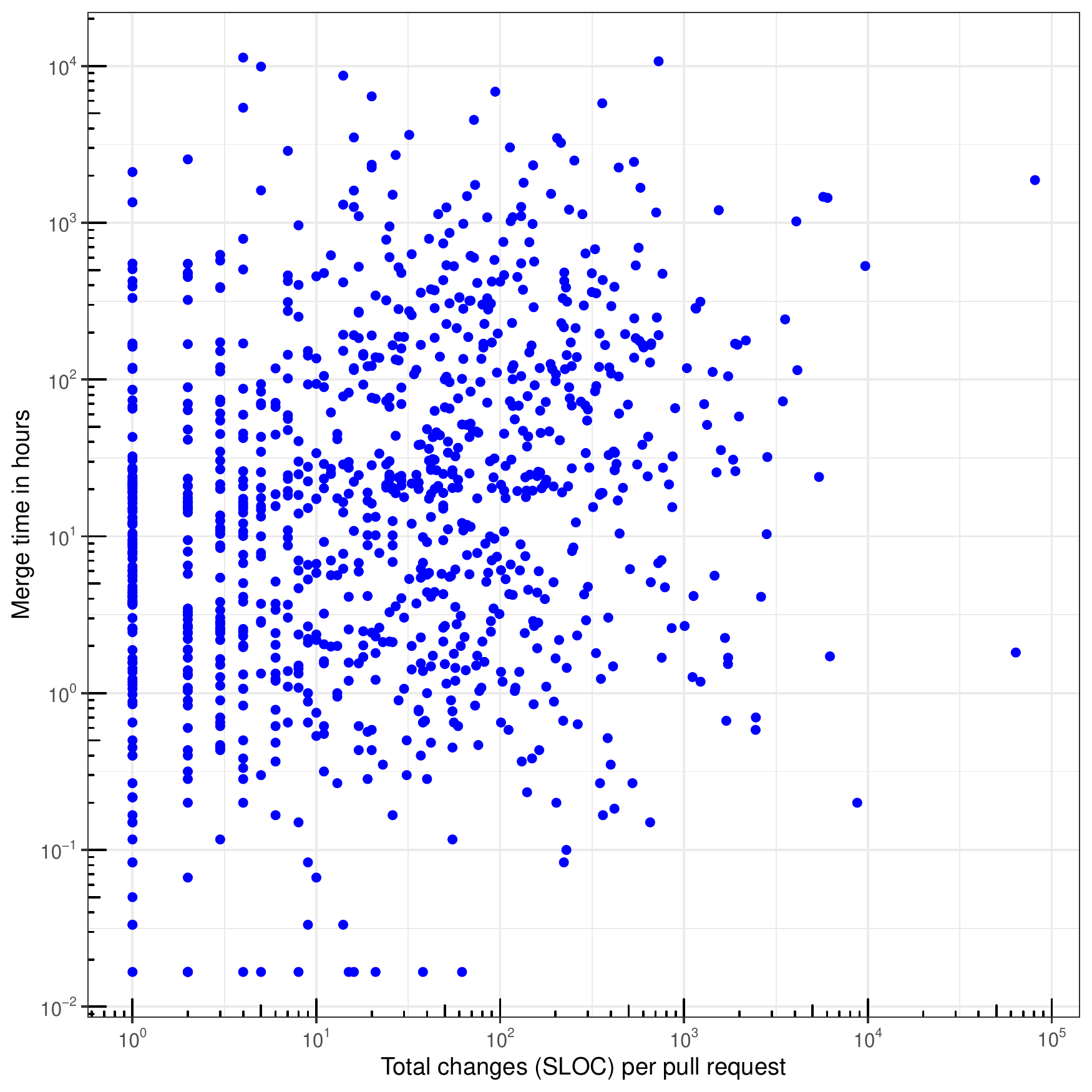}
    \caption{A scatter plot of \pr\ sizes and \ttm. Random sample of $N=\num{1000}$ \pr s.}
    \label{fig:scatterplot_pr_sloc}
\end{figure}

In \cref{subsec:related-work}, we presented many studies investigating the relations of \pr\ size to merge speed.
These studies have reached contradictory conclusions.
When we calculate the Spearman correlation coefficient
between \pr\ size and \ttm\ on our dataset,
the strength of the relation is weak ($r_s = 0.26, p < .001$).
Our interpretation of correlation coefficients relies on well-accepted guidance used in medicine and psychology~\cite{akoglu_2018,mukaka_guide_2012,schober_correlation_2018}.
We even looked at the scatter plot of \num{1000} randomly selected \pr s from our dataset for other patterns (see \cref{fig:scatterplot_pr_sloc}).
Both axes use a logarithmic scale for better display.
We observed no patterns and a lack
of a strong relationship between \pr\ size and \ttm.

Another factor that we expected to impact the \ttm\ was the ratio of different
change types to the total size of a \pr.
Intuitively, one can expect that the deletions or modifications of code would
be reviewed and merged faster than insertions of new code.
Therefore, the \pr s, which mainly  consist of insertions, should take
longer than those where most of the changes are deletions.

Our study shows that the strength of correlation coefficients between \ttm\ and ratios of different change
types is negligible
for insertions ($r_s=0.18, p < .001$),
deletions ($r_s=0.06, p < .001$),
and modifications ($r_s=-0.14, p < .001$).
One interesting observation is a minor negative
correlation between \ttm\ and the modification ratio, which is not present
for either insertion or deletion ratios.
As the ratio of modifications in a \pr\ increases,
the less time it takes to review that \pr, resulting in a 
reduced \ttm.
A possible explanation for this is that given the nature of the
modification as a change type
(existing code that has already been reviewed),
the reviewer assumes that the existing code is
correct and gives less scrutiny to the modifications.

\begin{boxed_new}
\textbf{RQ2:} \emph{We do not observe a strong 
relationship between the \ttm\ and the \pr\ size.
The finding is consistent across the various compositions of a \pr.}
\end{boxed_new}

\subsection*{RQ3: Does context influence the relationship of \pr\ size and composition to \ttm?}
\label{subsectionRQ3}

In RQ2, we found no strong relationship of the \pr\ size and composition to the \ttm.
In this section, we study whether different contexts influence
\pr\ size and composition, and hence \ttm. 

\subsubsection{Programming language}

\begin{table}[ht]
  \caption{Time-to-merge (hours) and \textsc{SLOC} per \pr. N = total number, M = mean, Mdn = median, SD = standard deviation.}
  \label{tab:after-outlier-data}
  \begin{tabular}{lrrrrrrr}
    \toprule
    &&\multicolumn{3}{c}{Time-to-merge}&
    \multicolumn{3}{c}{SLOC}\\
    \cmidrule(lr){3-5}
    \cmidrule(lr){6-8}
    Language&\textit{N}&\textit{M}&\textit{Mdn}&\textit{SD}&\textit{M}&\textit{Mdn}&\textit{SD}\\
    \midrule
    C& \num{45555} & 154& 15& 654& 313& 21& \num{5603}\\
    C++& \num{136308} & 190& 19& 769& 721& 21& \num{14574}\\
    C\#& \num{82387} & 161& 22& 639& 1416& 40& \num{23422}\\
    Java& \num{64495} & 173& 17& 886& 425& 43& \num{9169}\\
    JavaScript& \num{72510} & 140& 11& 654& 307& 15& \num{6904}\\
    PHP& \num{82074} & 194& 16& 809& 550& 15& \num{13795}\\
    Python& \num{137365} & 228& 21& 978& 229& 21& \num{7476}\\
    Ruby& \num{78814} & 183& 10& 961& 187& 13& \num{9096}\\
    Shell& \num{14212} & 545& 15& \num{2715}& 173& 8& \num{1811}\\
    TypeScript& \num{112539} & 122& 12& 523& 979& 35& \num{33703}\\
    \bottomrule
  \end{tabular}
\end{table}

\cref{tab:after-outlier-data} presents the descriptive statistics for
\textsc{SLOC} and \ttm\ for different programming languages.
The medians for both \ttm\ and \pr\ size vary by 2–-3 times.
A Kruskal-Wallis test for stochastic dominance~\cite{kruskal_use_1952} revealed that there was a statistically significant difference between the mean ranks of \ttm\ for at least one pair  of languages $(H (\num{9}) = \num{9788.44}, p < .001)$.
After performing a post hoc pairwise Dunn test~\cite{dunn_1961,dunn_1964} with a
Bonferroni correction~\cite{bonferroni} we observe a difference between
all languages ($p < .001$) except the following pairs:
C versus Java, C versus Shell, Java versus Shell, and JavaScript versus Ruby.

Similarly, a statistically significant difference between the mean ranks of \pr\ size was present between at least one pair of languages $(H (\num{9}) = \num{23475.83}, p < .001)$.
Based on a Dunn test with a Bonferroni correction we observe
a difference between all pairs of languages ($p < .001$) except the following pairs:
C versus C++, C versus Python, C++ versus Python, and C\# versus Java.
We notice that object-oriented and scripting languages such as C\#, Java,
and TypeScript tend to have larger median \pr\ sizes.
Interestingly, TypeScript (a typed superset
of JavaScript) has a noticeably larger mean \pr\ size than JavaScript.
A possible explanation for the larger \pr\ size for these languages
may be that a more evolved set of integrated development environments
(\textsc{IDE}) is available.
Using a good \textsc{IDE} enables engineers to generate more code.
On the other hand, Shell, which has the smallest median
\textsc{SLOC} value, is a more concise language and lacks  \textsc{IDE} support available to other programming
 languages~\cite{nanz_2015}.
The various types of code changes in a \pr\ are not significantly different among programming languages. 

We expected the \ttm\ to differ between programming languages.
\cref{section:rq} pointed out that the \textsc{SLOC}
to solve a problem can differ among programming languages.
Our assumption is that reviewing \pr s with a smaller size takes less time than reviewing larger \pr s.
There is a possibility that \pr s in some languages merge faster than others.
Therefore, by only looking at the entire dataset, we may be missing these details. 
To investigate the relationship between \pr\ size and \ttm\ in detail, we divide the \pr s by programming languages. 
\cref{tab:ttm-spearman-ratio-cor} presents the
Spearman correlation coefficient between the \ttm\ in hours and total changes for each 
programming language.
Given that we consider the same data points repeatedly, the p-values are adjusted using Benjamini–Yekutieli procedure~\cite{benjamini_yekutieli}.

\begin{table}[ht]
  \caption{Spearman correlation coefficient $r_s (p < .001)$ between \ttm\ (hours) and total \textsc{SLOC}. N = total number.}
  \label{tab:ttm-spearman-ratio-cor}
  \begin{tabular}{lrr}
    \toprule
    Language&\textit{N}&$r_s$\\
    \midrule
C& \num{45555} & $0.31$ \\
C++& \num{136308} & $0.21$ \\
C\#& \num{82387} & $0.25$ \\
Java& \num{64495} & $0.23$ \\
JavaScript& \num{72510} & $0.23$ \\
PHP& \num{82074} & $0.25$ \\
Python& \num{137365} & $0.37$ \\
Ruby& \num{78814} & $0.34$ \\
Shell& \num{14212} & $0.23$ \\
TypeScript& \num{112539} & $0.20$ \\
  \bottomrule
\end{tabular}
\end{table}

We anticipated that changes in low-level languages
such as {C} and {C++} would take longer to review and merge.
Mainly because of the amount of meticulous effort required in verifying
the correctness of basic frequent and rudimentary operations like handling strings
(e.g., checking arguments to \texttt{strncpy}), managing heap allocations
(e.g., handling either \texttt{NULL} returned by \texttt{malloc}, or
exceptions thrown by \texttt{new}), verifying the error handling of
return codes, etc.
Other languages under review use either built-in string types
or utilize garbage collection as a memory management paradigm.
These expectations were not confirmed.

\subsubsection{Affiliation}

We explored differences in the \ttm\ per \pr\ between repositories owned by companies (Apple, Facebook, Microsoft, and Square) and those that are not.
We hypothesized that engineers working as employees of those companies have different incentives than volunteer contributors (e.g., adherence to specific deadlines, metrics associated with active \pr s, and organizational pressure to \textquote{ship faster}).
These differences can cause engineers working in the industry to
approach the \pr \ review process more aggressively.

To test our hypothesis, we scanned our projects and identified a total of 12 projects where the GitHub repository is owned by one of the industrial organizations mentioned above, indicating a non-trivial amount of involvement by those companies. 
The \pr\ size for \gh\ repositories owned by industrial organizations ($Mdn$ = \num{43}) is higher than those that are not ($Mdn$ = \num{20}).
A Mann-Whitney \textit{U} test~\cite{mann_whitney} indicated that this difference was statistically significant $U(N_{Industry} = \num{100000}, N_{Non-industry} = \num{100000}) = \num{10923542062.5}, z = \num{71.58}, p < .001.$
The \ttm\ for \gh\ repositories owned by industrial organizations ($Mdn$ = \num{14.13}) is lower than those that are not ($Mdn$ = \num{17.43}).
A Mann-Whitney \textit{U} test indicated that this difference too was statistically significant $U(N_{Industry} = \num{100000}, N_{Non-industry} = \num{100000}) = \num{9727540456.5}, z = \num{-21.11}, p < .001.$
However, we do not see any significant difference in the correlation
coefficients between \ttm\ and total size of the \pr.
For industry projects ($r_s=0.25, p < .001$) and for non-industry projects ($r_s=0.27, p < .001$) the ratio of different types of code changes and their relationship
to \ttm\ is almost identical.

\subsubsection{Day of the week}

One potential issue with how we interpreted our data
was a failure to account for the weekend.
It would be reasonable to assume that most engineers do not constantly work over the weekends for typical industrial projects.
Such a practice would systematically introduce longer \ttm\ periods for \pr s submitted on a Friday versus Monday. 

However, we would argue that the context on GitHub is different from industrial projects, mitigating the impact of the weekend. Engineers working on a GitHub project may work in their spare time (evenings, holidays, weekends). Contributors are distributed around the world and therefore originate from different time zones. The working week may start and end on different days depending on the country. 

\begin{table}[ht]
  \caption{Spearman correlation coefficient $r_s (p < .001)$ between \ttm\ in hours and total \textsc{SLOC} in a \pr\ by \pr\ creation day. N = total number.}
  \label{tab:mtm-spearman-per-weekday}
  \begin{tabular}{lrr}
    \toprule
    Day created&\textit{N}&$r_s$\\
    \midrule
Monday& \num{132488} & $0.26$ \\
Tuesday& \num{145908} & $0.27$ \\
Wednesday& \num{146326} & $0.27$ \\
Thursday& \num{143626} & $0.27$ \\
Friday& \num{131587} & $0.27$ \\
Saturday& \num{65845} & $0.25$ \\
Sunday& \num{60479} & $0.28$ \\
  \bottomrule
\end{tabular}
\end{table}

We decided not to apply any special treatment to the \pr s submitted over the weekend for the above-mentioned reasons.
One observation we can make from \cref{tab:mtm-spearman-per-weekday}
is that the number of \pr s submitted over the weekend significantly 
drops.
Given the nature of the typical workweek, this finding is expected.
However, regardless of when a \pr\ was created~\cite{sliwerski_2005}, we still
do not see any relationship between the \ttm\ and \pr\ size.
Given that we consider the same data points repeatedly, the p-values are adjusted using Benjamini–Yekutieli procedure.
We also do not observe any significant difference in the ratio of
different types of code changes and median size of \pr s between
various days of week.
This finding suggests no particular period during
the week when engineers tend to make changes of a certain type or size.

\begin{boxed_new}
\textbf{RQ3:} \emph{The \textsc{SLOC} and \ttm\ differ between the programming languages. 
Industry projects tend to have shorter \ttm\ and contain more \textsc{SLOC} per \pr.
Day of the week when \pr\ was created influences the \pr\ count.}
\end{boxed_new}

\subsection*{Replication}

To determine whether our findings could be replicated across different datasets, we applied a subset of our analysis to projects using Gerrit and Phabricator.
Gerrit and Phabricator datasets have the added benefit of being able
to track both the \tta\ and \ttm.
Though the number of projects in Gerrit and Phabricator datasets 
is smaller (8 projects with \num{401790} code reviews)
than we extracted from \gh,
and more homogeneous (mainly in C or C++),
the results of the analysis are the same:
the size of the \pr\ and its composition has a weak correlation
to \tta\ and \ttm\ (refer to Table~\ref{tab:ttm_vs_tc_gerrit_and_phab_projects}).

\begin{table}[ht]
  \caption{Spearman correlation coefficient $r_s (p < .001)$ for \tta\ (\textsc{TTA}) and \ttm\ (\textsc{TTM}) in hours versus total changes. N = total number.}
  \label{tab:ttm_vs_tc_gerrit_and_phab_projects}
  \begin{tabular}{llrrr}
    \toprule
    Project&Review tool&\textit{N}&${r_s}_\textsc{TTA}$&${r_s}_\textsc{TTM}$\\
    \midrule
Eclipse &  Gerrit &\num{8962} & 0.25 & 0.24\\
GerritHub &  Gerrit &\num{31826} & 0.27 & 0.26\\
LibreOffice &  Gerrit &\num{17534} & 0.11 & 0.13\\
OpenStack &  Gerrit &\num{151743} & 0.28 & 0.28\\
Blender &  Phabricator &\num{5224} & 0.26 & 0.26\\
FreeBSD &  Phabricator &\num{15984} & 0.28 & 0.27\\
LLVM &  Phabricator &\num{71082} & 0.32 & 0.30\\
Mozilla &  Phabricator &\num{99435} & 0.28 & 0.30\\
    \bottomrule
  \end{tabular}
\end{table}

\section{Discussion}
\label{section:discussion}

\subsection{Interpretation of our results}

Regardless of how the \pr s are partitioned
(day of creation, language, and affiliation), the data shows no significant correlation between the size and composition of
\pr\ and \ttm.
The size of a \pr\ is not a factor that can be meaningfully changed to speed up code changes.
Engineers will need to set realistic expectations about how much
of an actual control they have over how fast their changes will be merged.
It is possible that social characteristics identified in earlier studies,
such as an engineer's experience or reputation, are more meaningful variables in improving code velocity~\cite{gousios_exploratory_2014,tsay_influence_2014}.

One unexpected finding is that no code change type merged faster than others.
Intuitively, one would expect that adding new code to the repository should take more time than removing or modifying the existing code.  There are many possibilities that
can explain the lack of differences in the \ttm\
\begin{enumerate*}[label=(\roman*),before=\unskip{ }, itemjoin={{, }}, itemjoin*={{, and }}]
    \item reviews of deletions or modifications are accepted faster,
    but the review process
    itself represents a minor part in the overall \ttm\ and, therefore,
    the differences are not noticeable
    \item reviewers lack sufficient domain knowledge about
    the history of the codebase and, therefore, require the same
    amount of time to inspect modifications and deletions as it would
    take for new code
    \item projects based on the repositories we mined do not treat
    removing obsolete code as a priority.
\end{enumerate*}

Our study shows that a programming language matters when it comes to
the \textquote{verbosity} of changes. We see that more
readable and expressive languages such as C\#, Java, and TypeScript
tend to have bigger \pr s.
In contrast, more concise languages such as various dialects of shell scripts have 2–3 times smaller median change size. 
While not related to \ttm, our study highlights the need 
to account for programming language when measuring an engineer's productivity~\cite{nguyen_2007}.

Evidence suggests that changes will be accepted faster on industry projects than non-industry projects and that the median \textsc{SLOC} per \pr\ is higher for industry projects. 
This finding seems reasonable because most industrial projects are based on a set of deadlines and are driven by various metrics such as an active bug count, code velocity, and outstanding pull request count.
In other words, engineers in the industry are incentivized
to merge incoming
\pr s faster. 
We noted in \cref{subsection:emp-code-size-findings}
that industry projects generally tend to have a bigger median size for the size of code changes. 

\subsection{Implications for research and practice}

\emph{Better understanding of the composition of \textsc{CI}/\textsc{CD} pipelines}.
Existing studies and recommendations about \pr\ size and composition are
focused on optimizing the code review acceptance time (\emph{\tta}).
However, there is not much publicly available data about the ratio of \tta\ to \ttm.
Therefore, it is unknown if focusing on reducing \tta\ versus other parts
of the \textsc{CI}/\textsc{CD} pipeline is the optimal way to reduce
the \ttm\ and thus increase the code velocity.
Getting more insight into the composition of the \textsc{CI}/\textsc{CD} pipeline
will help various organizations to determine what parts of the process
may need to be optimized and where the actual bottlenecks are.

\emph{Simplify the \pr\ size related guidance}.
In Observation~\ref{obs:median_code_change_differs}, we note that the existing
guidance is non-specific and varies by project. To avoid
misinterpreting subjective quantifiers such as \textquote{isolated}
and \textquote{small}, a project or organization may consider
specifying a number. An intriguing suggestion would be a number such as $7 \pm 2$ to ensure optimal
cognitive comprehension of code changes under review~\cite{miller_magical_1956}.
Previous research has shown that smaller changes are easier to
understand and solicit actionable feedback~\cite{bacchelli_2013}.
Utilizing automatic decomposition of code changes may also help avoid engineers manually partitioning their changes if reviewers consider them \textquote{too big}~\cite{barnett_helping_2015}.

\section{Threats to Validity}
\label{section:threats-to-validity}

Like any other study, the results we present in our paper are
subject to certain categories of threats.
In this section we enumerate the threats to construct, internal,
and external validity~\cite{shull_guide_2008}.

Using the number of forks as a proxy for the popularity of a GitHub repository creates an issue with \emph{construct validity}.
Like any metric, the number of forks per repository is pliable for manipulation (e.g., forks can be created in an automated manner, a number of forks can be inactive, and a repository is forked with no \pr s). We mitigated these concerns by picking the repositories with the largest number of \pr s and manually verifying that a repository is under active development.

One of the concerns for \emph{internal validity} is a potential presence of
methodological errors in the way analysis and collection of the \pr\ data
is perfomed.
There are various perils associated with mining \gh\ for
data~\cite{kalliamvakou_-depth_2016}.
We carry the following limitations: we trust GitHub data about the \pr\ being successfully merged (or not) without manually mining the commits in Git history. We rely on GitHub to adequately summarize all the commits included in the \pr\ and generate the correct diff describing them.

Threats to \emph{external validity} are related to the application of
our findings in other contexts. 
We did not analyze the development of widely used \textsc{OSS} projects
such as Linux, OpenBSD, and PostgreSQL, where the code reviews are conducted by submitting
patches to various mailing lists.
However, we confirmed the essence of our findings in \cref{subsectionRQ3}
with a subset of \textsc{OSS} projects where Gerrit or Phabricator code review data
was available.
Focusing on \textsc{OSS} code versus closed source code is another
issue in this category.
Though many organizations developing commercial software (e.g., Apple, Facebook, Google, Microsoft, and Twitter) have embraced the GitHub software development model for selected projects, most of their code is still developed in a closed-source model.
Therefore, we cannot verify that our findings are valid in a closed-source model.

\section{Conclusions and Future Work}
\label{section:conclusions-and-future-work}
This study presents whether the \pr\ size and composition can be meaningfully changed to increase code velocity.
We selected \num{100} popular, actively developed projects on \gh\ to study the relations of \pr\ size and composition to code velocity measured as \ttm.
Our analysis shows no relationship of \pr\ size and composition to the \ttm, regardless of how we partition the data:
day of the week the \pr\ was created,
affiliation to industry,
and programming language.
We found no patterns even though \pr s affiliated with the industry are larger and take less time than non-industry equivalents.
Our results remain the same on two other platforms: Gerrit and Phabricator, which also offered another indicator of code velocity---\tta.

Our finding that changing size or composition does not influence code velocity prompts new questions. For instance, 

\emph{Influence of personal traits and interpersonal relationships to \ttm}.
Other than \pr\ size and composition, engineers have limited control over their behavior, communication, and mannerism while interacting with other
team members. 
A study has shown that an engineer's reputation in the general
community and the project is a good predictor for \pr\ acceptance~\cite{baysal_influence_2013}.
Intuitively, it makes sense that engineers who have good communication
skills, are well-versed in conflict resolution, and display empathy
towards other participants will receive more cooperation.

\bibliographystyle{ACM-Reference-Format}
\bibliography{msr-2022}

\end{document}